\documentclass[conference]{IEEEtran}
\IEEEoverridecommandlockouts
\usepackage{cite}
\usepackage{amsmath,amssymb,amsfonts}
\usepackage{algorithmic}
\usepackage{graphicx, multirow}
\usepackage{textcomp}
\usepackage{xcolor}
\usepackage{url}

\def\BibTeX{{\rm B\kern-.05em{\sc i\kern-.025em b}\kern-.08em
    T\kern-.1667em\lower.7ex\hbox{E}\kern-.125emX}}
\begin{document}

\title{Deep Multilayer Perceptrons for Dimensional \\ Speech Emotion Recognition}
\author{\IEEEauthorblockN{Bagus Tris Atmaja}
\IEEEauthorblockA{\textit{School of Information Science} \\
\textit{Japan Advanced Institute of Science and Technology}\\
Nomi, Japan \\
bagus@jaist.ac.jp}
\and
\IEEEauthorblockN{Masato Akagi}
\IEEEauthorblockA{\textit{School of Information Science} \\
\textit{Japan Advanced Institute of Science and Technology}\\
Nomi, Japan \\
akagi@jaist.ac.jp}}

\maketitle

\begin{abstract}
Modern deep learning architectures are ordinarily performed on 
high-performance computing facilities due to the large size of the input features and 
complexity of its model. This paper proposes 
traditional multilayer perceptrons (MLP) with deep layers and small input size 
to tackle that computation requirement limitation. 
The result shows that our proposed deep MLP 
outperformed modern deep learning architectures, i.e.,  LSTM and CNN, on 
the same number of layers and value of parameters. The deep MLP exhibited the highest 
performance on both speaker-dependent and speaker-independent scenarios 
on IEMOCAP and MSP-IMPROV corpus. 
\end{abstract}


\section{Introduction}
Speech emotion recognition is currently gaining interest from both academia 
and commercial industries. Researchers in the affective computing field progressively 
proposed new methods to improve the accuracy of automatic emotion recognition. 
Commercial industries are trying to make this technology 
available to the market since its potential applications. Previously, 
researchers has attempted to implement speech-based emotion recognition for 
wellbeing detection \cite{Gao2018}, call center application \cite{Petrushin1998}, and automotive safety 
\cite{Nass2005}.

One of the common requirements in computing speech emotion recognition is the 
availability of high-performance computing since the dataset usually is very large 
in size, and the classifying methods are complicated. 
Graphical processing units (GPU)-based computers are often used 
over CPU-based computers due to its processing speed to process the data, particularly, 
image-like data. 

This paper proposes the use of deep multilayer perceptrons (MLP) to overcome
the requirement of high computing power required by modern deep learning
architectures. The inputs are high-level statistical functions (HSF), which are
used to reduce the dimension of input features. The outputs are emotion
dimensions, i.e., 
degree of valence, arousal, and dominance. 

According to research in psychology, dimensional emotion is another view 
in emotion theories apart from categorical emotion.
Russel \cite{Russell1979} argued that emotion categories could be derived from this 
dimensional emotion, particularly in valence-arousal space. 
Given the benefit of the ability to convert dimensional 
emotion to categorical emotion, but not vice versa, 
predicting the emotion dimension is more beneficial than 
predicting the emotion category. We added dominance, since it is suggested in 
\cite{Bakker2014}, and the availability of those labels in the datasets. 
Dimensional emotion recognition are evaluated with deep MLP 
from speech data since the target applications are speech-based technology like call center 
and voice assistant applications.

The contribution of this paper is the evaluation of the classical MLP technique 
with deep layers compared with modern deep learning techniques, i.e., LSTM and CNN, 
in terms of concordance correlation coefficient (CCC). 
The deep neural network is an extension of the neural network with deeper layers, 
commonly five or more layers \cite{Atmaja2019d}.
Our results show that on both speaker-dependent and speaker-independent,
in three datasets,  
deep MLP obtained higher performances than LSTM and CNN. 
The proposed method worked effectively on the small size 
of feature, in which, this may be a limitation of our proposed deep MLP method.

\section{Data and Feature Sets}
In this section, we describe data and feature sets used in this 
research. 
\paragraph{IEMOCAP} The interactive emotional dyadic motion capture database 
is used in this research \cite{Busso2008}. Although the database consists of the measurement of 
speech, facial expression, head, and movements of affective dyadic sessions, 
only speech data are processed. The database contains approximately 12 h of data  
with 10039 utterances. All of those data are used.
The dimensional emotion labels are given in 
valence, arousal, and dominance, in range [1-5] score and normalized 
into [-1, 1] as in \cite{parthasarathy2017jointly} when those labels are fed into the neural network system. 
For speech data, two versions are available in the dataset, stereo data per dialog 
and mono data per sentence (utterance). We used the mono version since it is easy to 
process with the labels. The sampling rate of the data was 16 kHz and 16-bit PCM.

We arranged the IEMOCAP dataset into two scenarios, speaker-dependent 
(SD) and speaker-independent. On speaker-independent, we split the dataset with 
ratio 80/20 for training/test set, while in speaker-independent, the last 
session, i.e., session five, is left for the test set (leave one session out, LOSO). 
The ratio of dataset partition in the speaker-independent scenario is similar 
to speaker-dependent split which is shown in Table \ref{tab:data-partition}.

\paragraph{MSP-IMPROV} We used the MSP-IMPROV corpus to generalize the 
impact of the evaluated methods. MSP-IMPROV is an acted corpus of dyadic 
interactions to study emotion perception. This dataset consists of speech and 
visual recording of 18 hours of affective dyadic sessions. Same as IEMOCAP dataset, 
we only used the speech data, with 8438 utterances. The same split ratio 
is used in speaker-dependent scenario while the last 
session six is used for test set in in speaker-independent scenario, 
with the same labels scale and normalization. 
While IEMOCAP labels are annotated by 
at least two subjects, these MSP-IMPROV labels are annotated by at least 
five annotators. The speech data were mono, 44 kHz, 
and 16-bit PCM.

Table \ref{tab:data-partition} shows the number of utterances allocated 
for each set partition for both speaker-dependent and speaker-independent, 
including MSP-IMPROV dataset. 

\paragraph{Mixed-corpus} In addition to the two datasets above, we mixed 
those two datasets to create a new category of dataset namely mixed-corpus. 
In mixed-corpus, we concatenated speaker-dependent from IEMOCAP with 
speaker-dependent from MSP-IMPROV for each, training, development 
and test sets. The same rules also applied for the speaker-independent scenario.

\begin{table}[htbp]
\caption{Number of utterances used in the datasets on each partition}
\begin{center}
\begin{tabular}{l c c c}
\hline
Scenarios   &    Training   & Development   & Test  \\
\hline
IEMOCAP-SD  &   6431   &    1608    &   2000\\
IEMOCAP-LOSO    &   6295    &   1574    &   2170\\
IMPROV-SD   &   5256    &   1314    &   1868\\
IMPROV-LOSO     &   5452    &   1364    &   1622\\
\hline
\end{tabular}
\label{tab:data-partition}
\end{center}
\end{table}

\paragraph{Acoustic Feature Set}
We used high statistical functions of the low-level descriptor (LLD) from Geneva Minimalistic Acoustic Parameter Set (GeMAPS), which is developed by Eyben et al. \cite{eyben}. The HSF features are extracted per utterance depend on the given labels, 
while the LLDs are processed on a frame-based level with 25 ms window size and 10 ms of hop size. The use of HSF feature reduce computation complexity since the feature size decrease from (3409 $\times$ 23) to  (1 $\times$ 23 features), that is, for 
IEMOCAP dataset. To obtain the HSF feature, however, LLDs must obtained first. Then, HSF can be calculated as statistics of those LLDs for 
a fixed time, in this case, per utterance.
\begin{table}
\centering
\caption{GeMAPS feature \cite{eyben} and its functionals; only functionals 
        (HSFs) used in this dimensional SER.}
\begin{tabular}{l p{6.5cm}}
\hline
LLDs & loudness, alpha ratio, hammarberg index, spectral slope 0-500 Hz, spectral slope 
500-1500 Hz, spectral flux, 4 MFCCs, F0, jitter, shimmer, Harmonics-to-Noise Ratio (HNR), 
Harmonic difference H1-H2, Harmonic difference H1-A3, F1, F1 bandwidth, F1 
amplitude, F2, F2 amplitude, F3, and F3 amplitude. \\
\hline
HSFs & mean (of LLDs), standard deviation (of LLDs), silence \\
\hline
\end{tabular}
\label{tab:feature}
\end{table}

To add those functionals, we used a silence feature, which is also extracted per utterance. 
Silence feature, in this paper, is defined as the portion of the silence frames compared 
to the total frames in an utterance. In human-human communication, this portion of 
silence in speaking depends on the speaker's emotion. For example, 
high arousal emotion category like happy may have fewer silences (or pauses) 
than a sad emotion category. The ratio of silence per utterance is calculated as

\begin{equation} \label{eq:silence}
    S = \frac{N_{s}}{N_{t}},
\end{equation}
where $N_s$ is the number of frames to be categorized as silence (silence frames), 
and $N_t$ is the number of total frames within an utterance. To be categorized as 
silence, a frame is checked whether it is less than a threshold, which is a 
multiplication of a factor with a root mean 
square (RMS) energy ($X_{rms}$). This RMS energy is formulated as 

\begin{equation}
    th = 0.3 \times \overline{X_{rms}}
\end{equation}
and $X_{rms}$ is defined as

\begin{equation}
    X_{rms} = \sqrt{\frac{1}{n}\sum_{i=1}^{n}x[i]^2}
\end{equation}
where silence factor of $0.3$ is obtained from experiments.
These equations are similar to what is proposed in \cite{Sahu} 
and \cite{moore2014word}. In \cite{Sahu}, the 
author of that paper used a fixed threshold, while we evaluated some factors of 
to find the best factor for silence feature in speech emotion 
recognition. In \cite{moore2014word}, the author divided the total duration of 
disfluency over the total utterance length on $n$ words and 
counted it as a disfluency feature. 

\section{Benchmarked and Proposed Method}
We evaluated three different methods in this paper, LSTM, CNN and MLP. 
LSTM and CNN are used as baselines, while MLP with deep layers is the proposed 
method. All evaluated methods used the same numbers of layers, units 
and value of parameters. 

\subsection{Benchmarked Methods: LSTM and CNN}
LSTM and CNN are two common deep learning architectures widely used in 
speech emotion recognition \cite{Xie2019, Schmitt2019, Atmaja2019}. 
We used those two architectures as the baselines due 
to its reported effectiveness on predicting valence, arousal, and dominance. 
Both LSTM and CNN evaluated here have the same five layers with 
the same number of units. For the size of the kernel in CNN architecture, we 
determined it in order that the number of trainable parameters is similar. 
The other parameters, like batch size, feature and label standardization, 
loss function, number of iterations, and callback criteria, are same 
for both architectures. 

\begin{figure}[htbp]
\centerline{\includegraphics[width=0.45\textwidth]{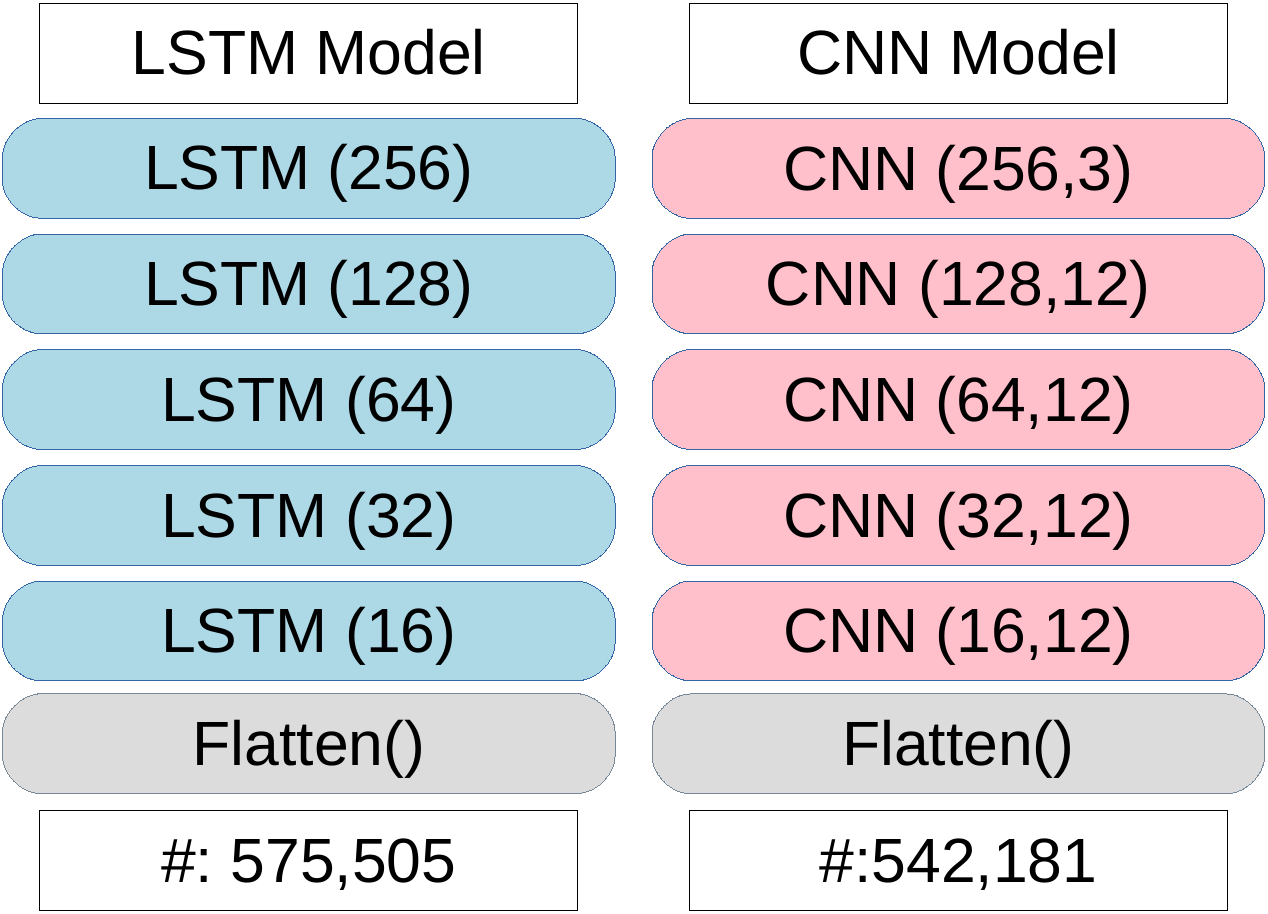}}
\caption{Diagram of LSTM and CNN used for benchmarking of proposed deep MLP; the number inside the 
bracket denoted the number of units (nodes); the second number on CNN denotes kernel size, 
\# denotes number of trainable parameters.}
\label{fig:lstm-cnn}
\end{figure}

Fig. \ref{fig:lstm-cnn} shows both structures of LSTM and CNN. On the 
first layer, 256 neurons are used and decreased half for the next layers since 
the models are supposed to learn better along with those layers. Five LSTM layers 
used tanh as activation function, while five CNN layers used ReLU activation 
function. We kept all output from the last LTM layer and flatten it before splitting 
into three dense layers for obtaining the prediction of valence, arousal, 
and dominance. For the CNN architecture, the same flatten layer 
is used before entering three one-unit dense layers.

Both architectures used the same standardization for input features and labels. 
The z-score normalization is used to standardize feature, while minmax scaler 
is used to scale the labels into [0, 1] range. We used CCC 
\cite{lawrence1989concordance} loss as the cost function 
with multitask (MTL) approach in which the prediction of valence, arousal, 
and dominance are done simultaneously. While CCC loss (CCCL) is used as the cost 
function, the following CCC is used to evaluate the performance of recognition. 
\begin{align}
CCC &= \dfrac{2\rho \sigma_x \sigma_y}{\sigma_x^2 + \sigma_y^2 + (\mu_x - \mu_y)^2} \\
CCCL &= 1 - CCC
\end{align}
where $\rho$ is Pearson's correlation between gold standard $y$ and and 
predicted score $x$, $\sigma$ is standard deviation, and $\mu$ is the mean score. 
The total loss ($L_T$) for three variables is then defined as the sum of CCCL for 
those three with corresponding weighting factors,
\begin{align}
L_T  = \alpha CCCL_V + \beta CCCL_A + (1-\alpha-\beta) CCCL_D
\end{align}
where $\alpha$ and $\beta$ are weighting factors for loss function of valence 
($CCCL_V$) and arousal ($CCCL_A$), respectively. The weighting factors of 
loss function for dominance ($CCCL_D$) is obtained by subtracting 1 by the sum 
of those weighting factors. The CCC is used to evaluate all architectures 
including the proposed deep MLP method.

All architectures used a mini-batch size of 200 utterances by shuffling 
its orders, maximum number iteration of 180, and 10 iterations patience 
of early stopping criteria (callback). 
An adam optimizer \cite{Kingma2014} is used to adjust the learning 
rate during the training process. Both architectures run on GPU-based
machine using CuDNN implementation 
\cite{Chetlur2014} within Keras toolkit \cite{chollet2015keras}.

\subsection{Proposed Method: Deep MLP}
\begin{figure}[htbp]
\centerline{\includegraphics[width=0.4\textwidth]{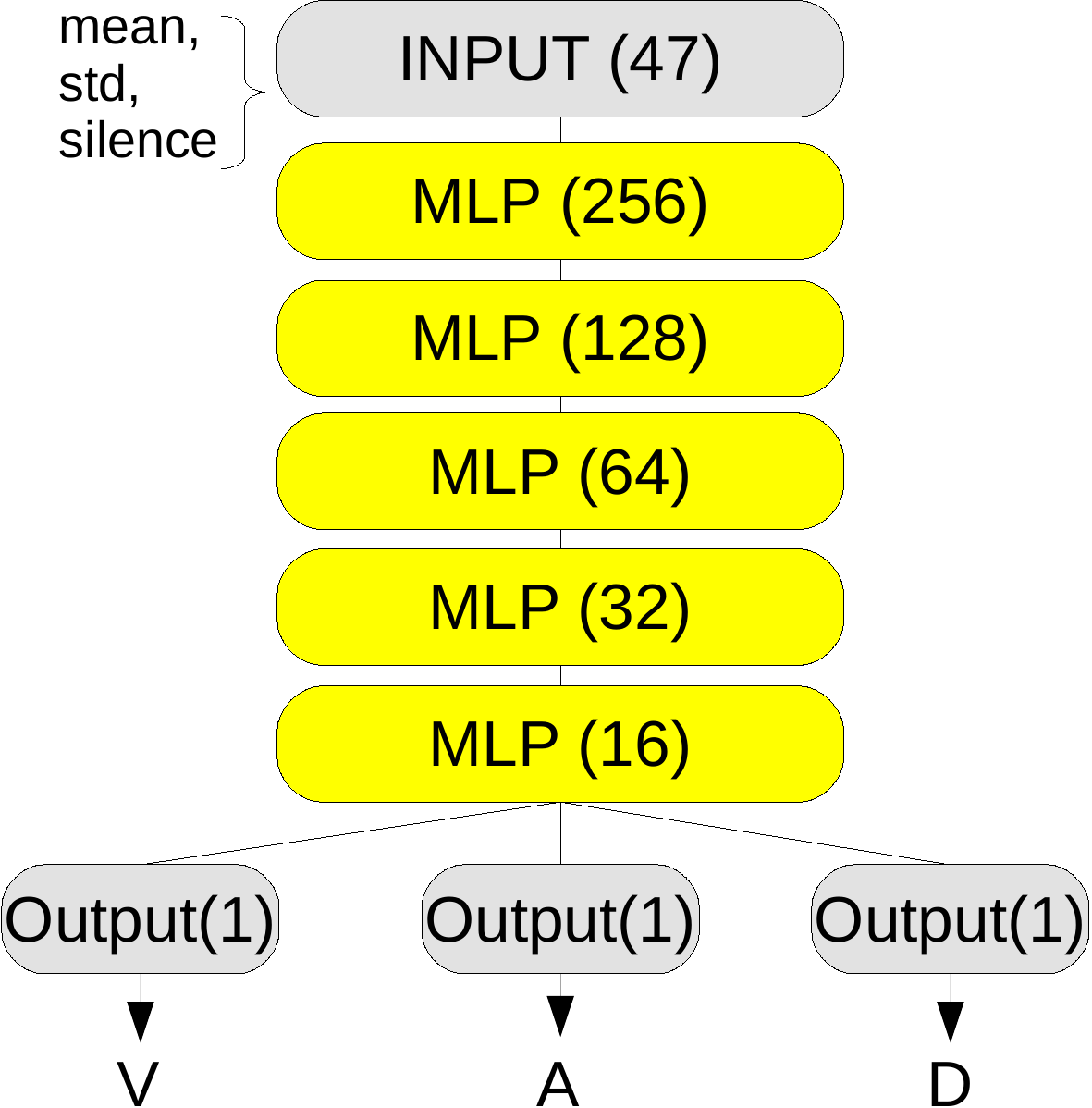}}
\caption{Diagram of proposed deep MLP with five layers; the number inside the 
bracket denoted the number of units.}
\label{fig:deepmlp}
\end{figure}

Fig. \ref{fig:deepmlp} shows our proposed deep MLP structure. The 
MLP used here similar to the definition of connectionist learning proposed 
by Hinton \cite{Hinton1989}. As the benchmarked methods, 
deep MLP also has five layers with the same number of units as previous. 
The only difference of the layer structure from the previous is the absent of 
flatten layer since the output of the last MLP layers can be coupled directly to 
three one-unit dense layers. While the previous LSTM and CNN used batch 
normalization layer in the beginning (input) to speed-up computation process, 
this deep MLP structure did not use that layer since the implementation 
only need a minute to run on a CPU-based machine.

We used the same batch size, tolerance for early stopping criteria, optimizer, 
and maximum number of iteration as the benchmarked methods. While the 
benchmarked methods used CCC the as loss function, the proposed deep MLP method used 
a mean square error (MSE) as the cost function, 
\begin{equation}
MSE = \dfrac{1}{n} \sum_{i=1}^n (y_i-x_i)^2.
\end{equation}
The total loss function is given as an average of MSE scores from valence, arousal, 
and dominance,
\begin{equation}
MSE_T = \dfrac{1}{3} (MSE_V + MSE_A + MSE_D).
\end{equation}
There are no weighting factors used here since we do not find a way to 
implement it via scikit-learn toolkit \cite{scikit-learn}, in which 
the proposed deep MLP is implemented. The same reason applied for the selection 
of MSE over CCC for the loss function. The Python implementation codes for 
both proposed and benchmarked methods are available at 
\url{https://github.com/bagustris/deep_mlp_ser}.

\section{Experiment Results and Discussion}
CCC is the standard metric used in affective computing to measure the 
performance of dimensional emotion recognition. We presented our results 
in that metric in two different groups; within-corpus and mixed-corpus 
evaluation. The results are shown in Table \ref{tab:result-within} and \ref{tab:result-mixed}. 

Table \ref{tab:result-within} shows CCC scores of valence (V), arousal (A), dominance (D)
and its average from different datasets, scenarios, and methods. The proposed 
deep MLP method outperforms benchmarked methods by remarkable margins. 
On every emotion dimensions and averaged score, the proposed deep MLP gained 
the highest CCC score for both speaker-dependent and speaker-independent 
scenarios (typed in bold). On IEMOCAP dataset, the score of speaker-dependent 
is only slightly 
higher than speaker-independent due to the nature of dataset structure. 
The utterances in IEMOCAP dataset is already in order by its session when it is 
sorted by file names. The change from speaker-dependent to speaker-independent is done by 
changing the number of train/test partitions. In contrast, the file naming 
of utterances in MSP-IMPROV made the arrangement of the sessions not in order 
when utterances are sorted by its file names. We did the sorting process 
to assure the pair of features and labels. The gap between 
speaker-dependent and speaker-independent in MSP-IMPROV is larger than in 
IEMOCAP which may be caused by those different files organization. A case where our deep MLP method gained a lower score 
is on dominance part of MSP-IMPROV speaker-dependent scenario, 
however, the averaged CCC score, in that case, is still the highest.

\begin{table}[htbp]
\caption{Results of CCC scores on within-corpus evaluation}
\begin{center}
\begin{tabular}{c l c c c c}
\hline
Dataset &   Method   &    V   & A   &   D   & Mean\\
\hline
\parbox[t]{2mm}{\multirow{8}{*}{\rotatebox[origin=c]{90}{IEMOCAP}}} 
& \multicolumn{5}{c}{\textit{speaker-dependent}} \\
& LSTM    &       0.222    &       0.508    &   0.432    & 0.387 \\
& CNN        &   0.086        &   0.433    &   0.401    & 0.307 \\
& MLP        &   \textbf{0.335}        &   \textbf{0.599}    &   \textbf{0.473}    & \textbf{0.469} \\
& \multicolumn{5}{c}{\textit{speaker-independent (LOSO)}} \\
& LSTM    & 0.210    & 0.474    & 0.394    & 0.359 \\
& CNN    & 0.113    & 0.460    & 0.410    & 0.328 \\
& MLP    & \textbf{0.316}    & \textbf{0.488}    & \textbf{0.454}    & \textbf{0.453} \\
\hline
\parbox[t]{2mm}{\multirow{8}{*}{\rotatebox[origin=c]{90}{MSP-IMPROV}}} 
& \multicolumn{5}{c}{\textit{speaker-dependent}} \\
& LSTM    & 0.392    & 0.629    & \textbf{0.524}    & 0.515 \\
& CNN    & 0.346    & 0.623    & 0.522    & 0.497 \\
& MLP    & \textbf{0.438}    & \textbf{0.650}    & 0.519    & \textbf{0.536} \\
& \multicolumn{5}{c}{\textit{speaker-independent (LOSO)}} \\
& LSTM    & 0.210    & 0.483    & 0.313    & 0.335 \\
& CNN    & 0.216    & 0.524    & 0.387    & 0.375 \\
& MLP    & \textbf{0.269}    & \textbf{0.551}    & \textbf{0.401}    & \textbf{0.407} \\
\hline
\end{tabular}
\label{tab:result-within}
\end{center}
\end{table}

Table \ref{tab:result-mixed} shows the results from the mixed-corpus dataset. This 
corpus is concatenation of IEMOCAP with MSP-IMPROV as listed in Table 
\ref{tab:data-partition}, for both speaker-dependent and speaker-independent 
scenarios. In this mixed-corpus, the proposed deep MLP method also outperformed 
LSTM and CNN in all emotion dimensions and averaged CCC scores. The score 
on speaker-dependent in that mixed-corpus is in between the score of 
speaker-dependent in IEMOCAP and MSP-IMPROV within-corpus. For 
speaker-independent, the score is lower than in within-corpus. This low score 
suggested that speaker variability (in different sessions) affected the 
result, even with the z-normalization process. Instead of predicting 
one different session (LOSO), the test set in the mixed-corpus consists 
of two different sessions, each from IEMOCAP and MSP-IMPROV, which made 
regression task more difficult. 

We showed that our proposed deep MLP functioned to overcome 
the requirement of modern neural network architectures since it surpassed 
the results obtained by those architectures. Using a small dimension of feature size, 
i.e., 47-dimensional data, our deep MLP with five layers, excluding input 
and output layers, achieved the highest performance. Modern deep learning 
architectures require high computation hardware, e.g., GPU card, which 
costs expensive. We showed that using a small deep MLP architecture, which 
does not require high computation load, better performance can be achieved. 
Our proposed deep MLP method gained a higher performance than benchmarked methods 
not only on both within-corpus and mixed-corpus but also on both 
speaker-dependent and speaker-independent scenarios. Although the proposed 
method used the different loss function from the benchmarked methods, 
i.e., MSE versus CCC loss, we presumed that our proposed deep MLP will 
achieve higher performance if it used the CCC loss since the evaluation 
metric is CCC.

\begin{table}[htbp]
\caption{Results of CCC scores on mixed-corpus evaluation}
\begin{center}
\begin{tabular}{l c c c c}
\hline
Method   &    V   & A   &   D   & Mean\\
\hline
\multicolumn{5}{c}{\textit{speaker-dependent}} \\
LSTM    &   0.262    &   0.518    &   0.424    &   0.401 \\ 
CNN    &   0.198    &   0.494    &   0.424    &   0.372 \\ 
MLP    &   \textbf{0.395}    &   \textbf{0.640}    &   \textbf{0.461}    &   \textbf{0.499} \\ 
\hline
\multicolumn{5}{c}{\textit{speaker-independent (LOSO)}} \\
LSTM    &   0.118    &   0.270    &   0.242    &   0.210 \\ 
CNN    &   0.073    &   0.265    &   0.249    &   0.196 \\ 
MLP    &   \textbf{0.212}    &   \textbf{0.402}    &   \textbf{0.269}    &   \textbf{0.294} \\ 
\hline
\end{tabular}
\label{tab:result-mixed}
\end{center}
\end{table}

\section{Conclusions}
This paper demonstrated that the use of deep MLP with proper parameter choices 
outperformed the more modern neural network architectures with the same number 
of layers. For both speaker-dependent and speaker-independent, the proposed 
deep MLP gained the consistent highest performance among evaluated methods. 
The proposed deep MLP also gained the highest score on both within-corpus and mixed-corpus 
scenarios. Based on the results of these investigations, there is no high requirements 
on computing power to obtain outstanding results on dimensional 
speech emotion recognition. The proper choice of 
feature (i.e., using small size feature) and the classifier can leverage the 
performance of conventional neural networks. 
Future research can be directed to 
investigate the performance of the proposed method on cross-corpus evaluations, 
which is not evaluated in this paper.

\bibliographystyle{IEEEtran}
\bibliography{deepmlp}

\end{document}